\UseRawInputEncoding
\documentclass[aps,10pt,prr,twocolumn,superscriptaddress,floatfix,nofootinbib,longbibliography]{revtex4-2}

\usepackage{graphicx,amsfonts,amssymb,amsmath,hyperref,enumerate}

\usepackage{algorithm}
\usepackage{algpseudocode}
\usepackage{pgfplots}
\usepackage{pgf}
\usepackage{lmodern}
\usepackage{import}
\usepackage{xr}
\usepackage{enumitem}
\usepackage{soul}
\usepackage[normalem]{ulem}
\usepackage{multirow}
\usepackage{flowchart}
\usetikzlibrary{arrows}
\usepackage{ulem}

\newif\ifhyper
\hypertrue
\ifhyper
\hypersetup{
   citecolor = {red},
   colorlinks = {true}, 
   urlcolor = {blue} 
}
\fi

\newcommand{\beq}{\begin{equation}}
\newcommand{\eeq}{\end{equation}}
\newcommand{\beqa}{\begin{eqnarray}}
\newcommand{\eeqa}{\end{eqnarray}}

\newcommand{\comment}[1]{}

\def\Longarrow{\protect\@lra}
\def\@lra{\relbar\joinrel\relbar\joinrel\relbar\joinrel\relbar\joinrel\rightarrow}

\pgfplotsset{compat=1.18}

\usepackage[caption=false]{subfig}
\usepackage{ragged2e}
\usepackage{etoolbox}

\begin{document} 

\title{Classical Neural Networks on Quantum Devices via Tensor Network Disentanglers:\\ A Case Study in Image Classification}

\author{Borja Aizpurua}

\affiliation{Multiverse Computing, San Sebasti\'{a}n, Spain}
\affiliation{Department of Basic Sciences, Tecnun - University of Navarra, San Sebasti\'an, Spain}

\author{Sukhbinder Singh}
\affiliation{Multiverse Computing, Toronto, Ontario, Canada}

\author{Rom\'{a}n Or\'{u}s}

\affiliation{Multiverse Computing, San Sebasti\'{a}n, Spain}

\affiliation{Donostia International Physics Center, San Sebasti\'an, Spain}

\affiliation{Ikerbasque Foundation for Science, Bilbao, Spain}

\begin{abstract}
We address the problem of implementing bottleneck layers from classical pre-trained neural networks on a quantum computer, with the goal of exploring intrinsically quantum ansatz for representing large linear layers within hybrid classical--quantum models. Our approach begins with a compression step in which the target linear layer is represented as an effective matrix product operator (MPO) without degrading model performance. The MPO is then further disentangled into a more compact form. This enables a hybrid classical-quantum execution scheme, where the disentangling circuits are deployed on a quantum computer while the remainder of the network---including the disentangled MPO---runs on classical hardware. We introduce two complementary algorithms for MPO disentangling: (i) an \emph{explicitly} disentangling variational method leveraging standard tensor-network optimization techniques, and (ii) an \emph{implicitly} disentangling gradient-descent-based approach. We validate these methods through a proof-of-concept translation of simple classical neural networks for MNIST and CIFAR-10 image classification into a hybrid classical-quantum form.
\end{abstract}

\maketitle

\section{Introduction}

The rapid improvement of quantum hardware has intensified the search for useful applications on present-day and near-term quantum devices. In parallel, advances in artificial intelligence have motivated growing interest in leveraging quantum computation to enhance modern machine learning methodologies. Research at this intersection has largely followed two complementary directions \cite{quantumAI}. The first develops hybrid classical-quantum algorithms---including the Variational Quantum Eigensolver (VQE) \cite{vqe, vqereview}, the Quantum Approximate Optimization Algorithm (QAOA) \cite{qoqa, qoqareview}, and quantum variants of classical ML methods such as Quantum Boltzmann Machines \cite{quantumboltzmann} and Variational Quantum Classifiers \cite{variationalclassifier}. The second explores purely quantum machine learning algorithms, such as quantum k-means clustering \cite{quantumclustering}, quantum principal component analysis \cite{quantumpca}, the Quantum Linear Systems Algorithm \cite{quantumlinearequations}, and Quantum Circuit Born Machines \cite{quantumbornmachine}.

Both approaches face significant limitations. In hybrid models, quantum circuits are often too shallow to capture meaningful structure \cite{holmes2022connecting, sim2019expressibility}, or too deep to train effectively on noisy hardware \cite{preskill2018quantum}. They may also lack the inductive biases crucial for specific ML tasks, and even when appropriately scaled, training can be hindered by barren plateaus \cite{cerezo2021variational, barrenplateau}. Purely quantum methods encounter similar scalability challenges, with current demonstrations restricted to small problem sizes.

In this work, we investigate how quantum computation can be integrated into practical, large-scale classical ML workflows. Specifically, we focus on embedding a bottleneck segment of a classically pre-trained deep neural network into a quantum circuit, with the aim of improving performance or efficiency. Modern deep networks concentrate much of their parameter count in large linear layers, each represented by a weight matrix. Concretely, given a network $\mathcal{N}$, we target a large bottleneck linear layer represented by matrix $W$ with dimension $2^\ell, \ell \in \mathbb{N}$, and implement it on a quantum processor by means of quantum circuits for $\ell$ qubits.

Inference proceeds in a hybrid fashion: classical data flows through the neural network until reaching the designated quantum layer. The output activations of the preceding classical layer are encoded into a quantum state---via amplitude encoding or by approximating them as a matrix product state (MPS) in canonical form, which is prepared on the quantum computer \cite{Ran_2020}---and processed by the pre-trained quantum circuit. The circuit's outputs are measured and converted back into classical activations through measurement and classical post-processing (full state tomography in our proof-of-concept simulations), and passed to the remainder of the classical network $\mathcal{N}$ for standard inference.


The central challenge in our approach is to translate the weight matrix $W$ into a quantum circuit of manageable complexity. A straightforward method involves selecting a circuit ansatz and optimizing it---via classical variational or gradient descent techniques---to approximate $W$. However, as discussed below, this optimization can become prohibitively expensive, with costs scaling exponentially in circuit depth, especially when $W$ is represented as a dense matrix.  

To address this, we propose a two-step strategy to better manage computational overhead. First, we perform a \textit{weight compression} step by approximating $W$ as a \textit{Matrix Product Operator} (MPO) \(M_\chi\), where \(\chi\) denotes the maximum bond dimension in the tensor network. Crucially, this replacement preserves the network's performance within acceptable bounds. MPO-based compression of linear layers has been demonstrated as an effective model compression technique \cite{tnn, tnnreview}, including applications in contemporary large language models \cite{tomut2024compactifai}.  

Having obtained the optimized MPO \(M\) approximating \(W\), we then apply a \textit{disentangling} step: we approximate \(M\) by quantum circuits such that  
\[
M \approx \mathcal{Q}_L \, M'_{\chi'} \, \mathcal{Q}_R,
\]  
where \(M'\) is a more compact MPO with reduced bond dimension \(\chi' < \chi\), and \(\mathcal{Q}_L, \mathcal{Q}_R\) are quantum circuits. We envision these circuits, transpiled to hardware-level implementations, running on a quantum processor, while the compressed MPO \(M'\) remains integrated within the classical neural network.  

We emphasize that the primary aim of this paper is not to demonstrate a practical quantum advantage for inference on deep learning models, but to initiate the development of concrete algorithms realizing this hybrid classical-quantum framework---one that could eventually enable new scaling regimes beyond what is accessible with purely classical tensor-network representations. Here, we demonstrate a compression technique in which classical parameters in a pretrained model are effectively and efficiently replaced---via a disentangling procedure---with parameters encoded in a quantum circuit. This, in turn, suggests that incorporating quantum circuits into a classical model can enhance its expressivity.


Accordingly, we do not claim quantum advantage in computational cost or inference speed, but rather study expressivity as a distinct scaling resource enabled by intrinsically quantum circuit ansatz.

Recently, we outlined this proposal in the context of large language models \cite{aizpurua2024quantum}. Here, we focus on a concrete proof-of-concept study involving image classification on the MNIST~\cite{deng2012mnist, lecun1998gradient} and CIFAR-10~\cite{krizhevsky2009learning} datasets, using elementary neural architectures such as multi-layer perceptrons (MLPs) and their MPO-tensorized counterparts. Even in this simplified setting, we identify significant challenges en route to practical quantum advantage---mainly, the rapid growth in circuit depth after transpilation to hardware---and propose several avenues for improving the baseline methods presented here.  

\begin{figure}[t]
    \centering
    \includegraphics[width=\linewidth]{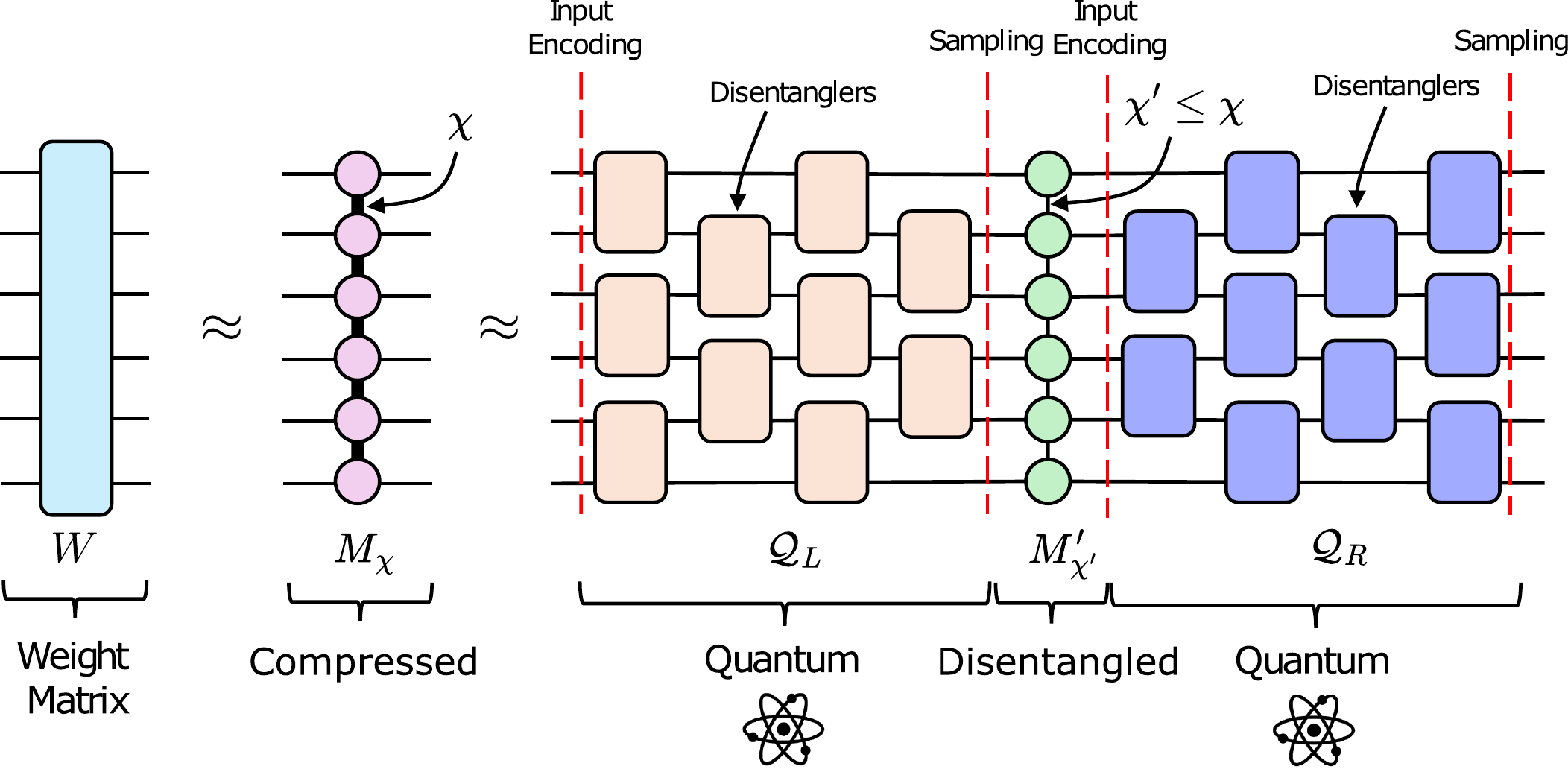}
    \caption{[Color online] Schematic of the hybrid quantum-classical model architecture. A classical image is encoded into a quantum state and transformed via a disentangling circuit $\mathcal{Q}_L$. A compressed MPO $M'_{\chi'}$ acts on this state, followed by another quantum circuit $\mathcal{Q}_R$. Both $\mathcal{Q}_L$ and $\mathcal{Q}_R$ are implemented as variational quantum circuits composed of two-qubit gates. The final state is sampled or reconstructed to yield a classical vector, completing the inference pipeline.}
    \label{fig:quantum_embedding}
\end{figure}

\section{MPO compression of the weight matrix}

The first step is to compress the weight matrix \(W\) into an MPO \(M\). While other tensor network architectures are possible, we focus on MPOs due to their demonstrated effectiveness in compressing large neural networks \cite{tnn, tnnreview}, including large language models (LLMs) \cite{tomut2024compactifai}. It is important to note that MPO \(M\) is not simply a low-rank approximation of \(W\). Empirically, typical pre-trained weight matrices exhibit near-maximal MPO bond dimensions, as expected: gradient descent optimization---the standard training procedure---does not inherently bias weights toward low-rank subspaces unless such bias is explicitly enforced by the architecture or training algorithm.

Indeed, we find that naively replacing the weight matrix with a low-rank MPO approximation significantly degrades model performance. Instead, after substituting \(W\) with a low-rank MPO approximation, the updated model must be \textit{healed}---namely, retrained or fine-tuned---to recover its original accuracy. Once the training loss converges and the model is sufficiently healed, we obtain the optimal MPO \(M\) corresponding to the original weight \(W\). With \(M\) fixed, we can then proceed to the disentangling step. However, before doing so, it is necessary to select an ansatz for the disentangling circuits.

\section{Fixing a disentangling circuit ansatz}  

Several design choices must be made to fix the structure of the disentangling circuits before gate optimization. Ideally, we seek to completely disentangle the MPO, achieving \(\chi' = 1\). In general, this corresponds to a disentangling quantum channel (superoperator) \(\Phi\) defined via \(\Phi(MM^\dagger) = \bigotimes_i M^{\circ}_i\), where \(\{M^{\circ}_i\}\) are positive matrices forming a completely disentangled MPO \(M^{\circ}\) (or possibly its square root).  

From the Kraus decomposition \cite{nielsen2010quantum}
\begin{equation}\label{eq:kraus}
\Phi(\cdot) = \sum_k U_k (\cdot) U_k^\dagger,
\end{equation}
we recover a set of disentangling unitaries \(\{U_k\}\). Each \(U_k\) can be approximated by a quantum circuit \(\mathcal{U}_k\), yielding an ensemble of disentangling circuits. In this work, we restrict attention to a single quantum circuit in the Kraus ensemble by fixing \(k=1\) and setting \(\mathcal{Q}_L = \mathcal{U}_1\). We further allow for a distinct adjoint circuit \(\mathcal{Q}_R \neq \mathcal{Q}_L^\dagger\).  

Following these choices, the geometry, sparsity, and gate set of the disentangling circuits must be specified. We primarily adopt a two-body brickwall architecture, possibly interspersed with layers of single-qubit gates. (Some of our experiments to draw comparisons extend to higher-body gates, up to six-body interactions.) 

For simplicity and numerical stability, we restrict ourselves to real orthogonal disentanglers in this work. This restriction is a deliberate modeling choice rather than a technical limitation: while modern automatic differentiation frameworks support complex-valued parameters and extending our approach to complex unitaries is straightforward, real-valued circuits define a smaller variational class that trades expressivity for potentially improving numerical stability and reducing contraction cost. For instance, recent results show that real-valued tensor networks composed predominantly of non-negative entries can be contracted much more efficiently than generic real-valued tensor networks, demonstrating that the tensor data type can significantly affect contraction cost~\cite{tnpositivebias}. We expect complex-valued disentanglers to increase expressivity in general, but leave a systematic comparison between real and complex circuits to future work.

A key practical consideration in choosing the circuit ansatz is minimizing the size and complexity of the \emph{physical} circuits resulting from transpilation to target quantum hardware. Shallow logical circuits composed of two-body random unitaries often transpile into very deep physical circuits with potentially non-local gate placements, thereby increasing execution costs on quantum devices. To mitigate this, we \emph{fix} some gates in the disentangling circuits to elementary gates native to the hardware. Specifically, in this work, we fix some or all two-body gates to CNOT gates. This approach potentially sacrifices some disentangling power but significantly reduces transpilation complexity.  

As an extreme choice, we fix all two-body disentanglers to CNOTs arranged in a brickwall pattern and optimize only the single-qubit gates to maximize disentangling. Remarkably, as we demonstrate, this constrained ansatz still achieves strong model performance, indicating that variational single-qubit gates applied on a frozen scaffold of CNOTs provide substantial disentangling capability while lowering transpilation overhead.  

\begin{figure}[t]
    \centering
    \includegraphics[width=0.95\linewidth]{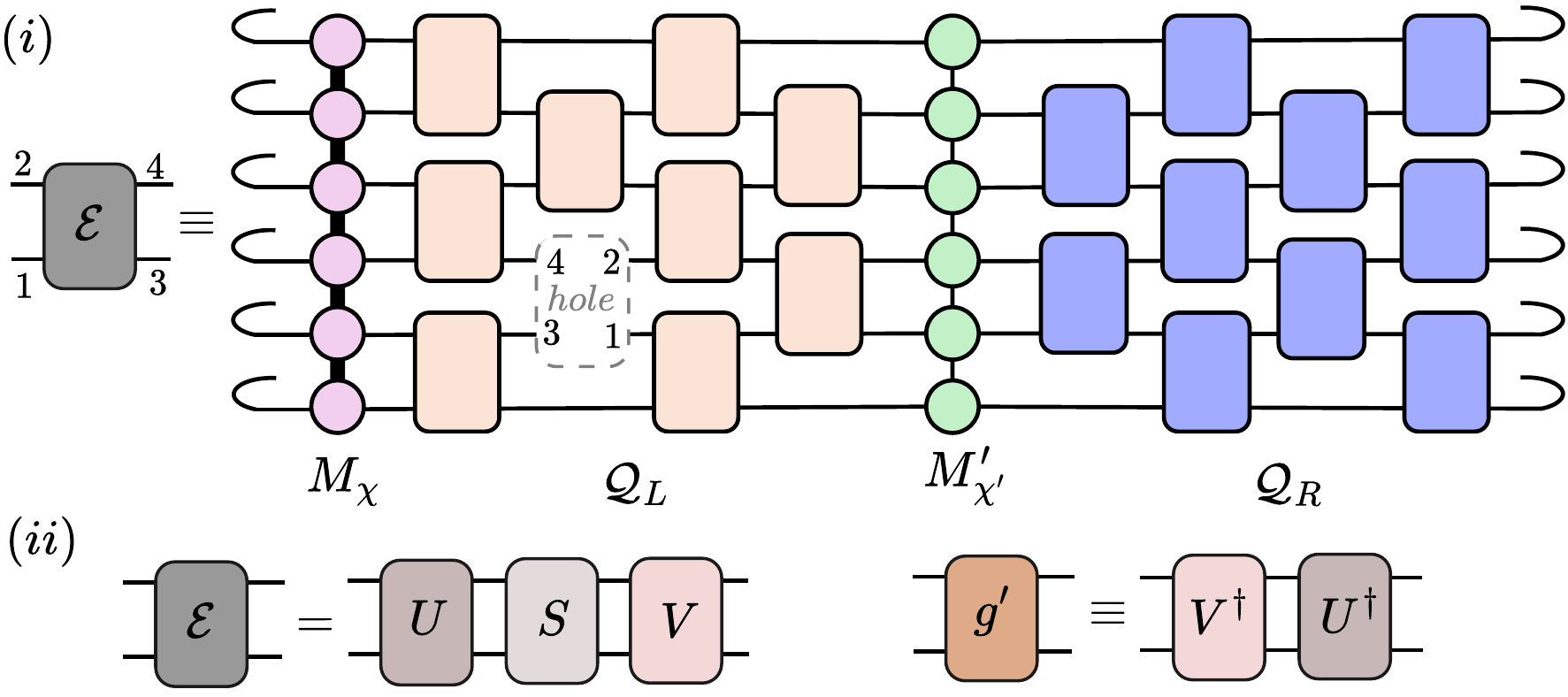}
    \caption{[Color online] Variational optimization of disentanglers. (i) The \emph{environment tensor} $\mathcal{E}_g$, Eq.~\ref{eq:env}, corresponding to a gate $g$ being optimized is obtained by removing that tensor from the circuit, leaving a hole with open indices, and contracting the remaining tensors that appear in Eq.~\ref{eq:env}. (ii) The singular value decomposition of $\mathcal{E}_g$ is performed, \(\mathcal{E}_g = U S V\). The updated gate is obtained as $g' = V^\dagger U^\dagger$.}
    \label{fig:disentangler_optimization}
\end{figure}

\section{Explicit Disentanglers via variational optimization}
One approach to find the disentanglers by maximizing the overlap  
\begin{equation}\label{eq:var_loss}
\mathrm{Tr}\left(M_\chi (\mathcal{Q}_L M'_{\chi'} \mathcal{Q}_R)\right) \big/ \left(\|M_\chi\| \, \|M'_{\chi'}\|\right),
\end{equation}
similar to technique used to determine disentanglers in the Multi-scale Entanglement Renormalization Ansatz (MERA) \cite{vidal2007entanglement, evenbly2009algorithms}.   

All disentangling gates are initialized randomly and then iteratively optimized. At each iteration, we compute the environment tensor \(\mathcal{E}_g\) for each gate \(g\). Concretely, assuming \(g\) belongs to the left circuit \(\mathcal{Q}_L\), the environment is given by  
\begin{equation}\label{eq:env}
\mathcal{E}_g \equiv \mathrm{Tr}\left(M_\chi \big(\mathcal{Q}^{g \rightarrow \mathrm{hole}}_L M'_{\chi'} \mathcal{Q}_R\big)\right) \big/ \left(\|M_\chi\| \, \|M'_{\chi'}\|\right),
\end{equation}
where \(\mathcal{Q}^{g \rightarrow \mathrm{hole}}_L\) denotes the circuit obtained by replacing gate \(g\) with an empty hole (i.e., a slot inside the quantum circuit with open indices, where $g$ plugs in), as illustrated in Fig.~\ref{fig:disentangler_optimization}. 

We note that the environment tensor $\mathcal{E}_g$ plays the role of the local linear functional governing the variation of the overlap objective with respect to the entries of the gate $g$. In this sense, $\mathcal{E}_g$ is closely related to the gradient of the cost function with respect to $g$, and for the circuit sizes considered in our proof-of-concept experiments, it could equivalently be obtained via automatic differentiation (AD) through the full contraction graph. However, the use of AD in this context, does not circumvent the dominant computational cost of contracting the full tensor network defining the overlap objective (as discussed below). Nonetheless, we also explored gradient-based explicit disentangling and found that environment-based variational updates converge faster and more robustly, as they exploit the local structure of the optimization problem rather than relying on repeated full-network gradient evaluations.

Computing the environments \(\mathcal{E}_g\) requires contracting the tensor network formed by the circuits and the MPOs, equating to the trace on the right hand side of Eq.~\ref{eq:env}. The gate \(g\) is then updated via singular value decomposition (SVD) of its environment, \(\mathcal{E}_g = U S V\), by setting  
\begin{equation}\label{eq:update}
g' \equiv V^\dagger U^\dagger.
\end{equation} 

In practice, we find that large weight matrices from pre-trained networks often translate into very deep circuits under this approach. Consequently, the cost of computing the \emph{exact} environment tensors \(\{\mathcal{E}_g\}\) grows rapidly with circuit depth, making the variational optimization computationally challenging for large circuits.

For the shallow circuits used in the numerical experiments presented here (in particular, MPOs defined over 6 qubits and modest circuit depth), exact contraction of the environment tensors is fully feasible and does not pose a computational bottleneck. Our discussion of approximate contraction strategies is included to clarify how the method would scale to deeper disentangling circuits, where contraction cost is governed primarily by circuit depth and entangling structure rather than the number of qubits alone.


However, for large circuits, these environments can be \emph{approximated} using standard tensor network techniques such as boundary MPS contractions or renormalization group methods \cite{tncontractionbook, tncontractiongray}. However, the efficiency of these approaches depends heavily on the circuit's structure, sparsity, and entangling power.  

We note that a similar variational procedure could be applied directly to approximate the dense weight matrix \(W\) using quantum circuits. For example, one could first decompose \(W\) via a QR decomposition, \(W = QR\), where \(Q\) is unitary, or via a singular value decomposition, \(W = USV\), where \(U\) and \(V\) are unitary matrices. The variational optimization would then target only the unitary components \(Q\) or \(U, V\), while the non-unitary components remain in the classical network.

Our motivation for not adopting this direct approach is that we expect the disentangling circuits for MPOs with small bond dimension---those typically obtained after the weight-compression step---to be substantially less entangling and more structured than those derived directly from dense matrices. Consequently, by first compressing the dense weight into an MPO, we anticipate a significant reduction in the overall computational cost.

\begin{figure}[t]
    \centering
    \includegraphics[width=0.55\linewidth]{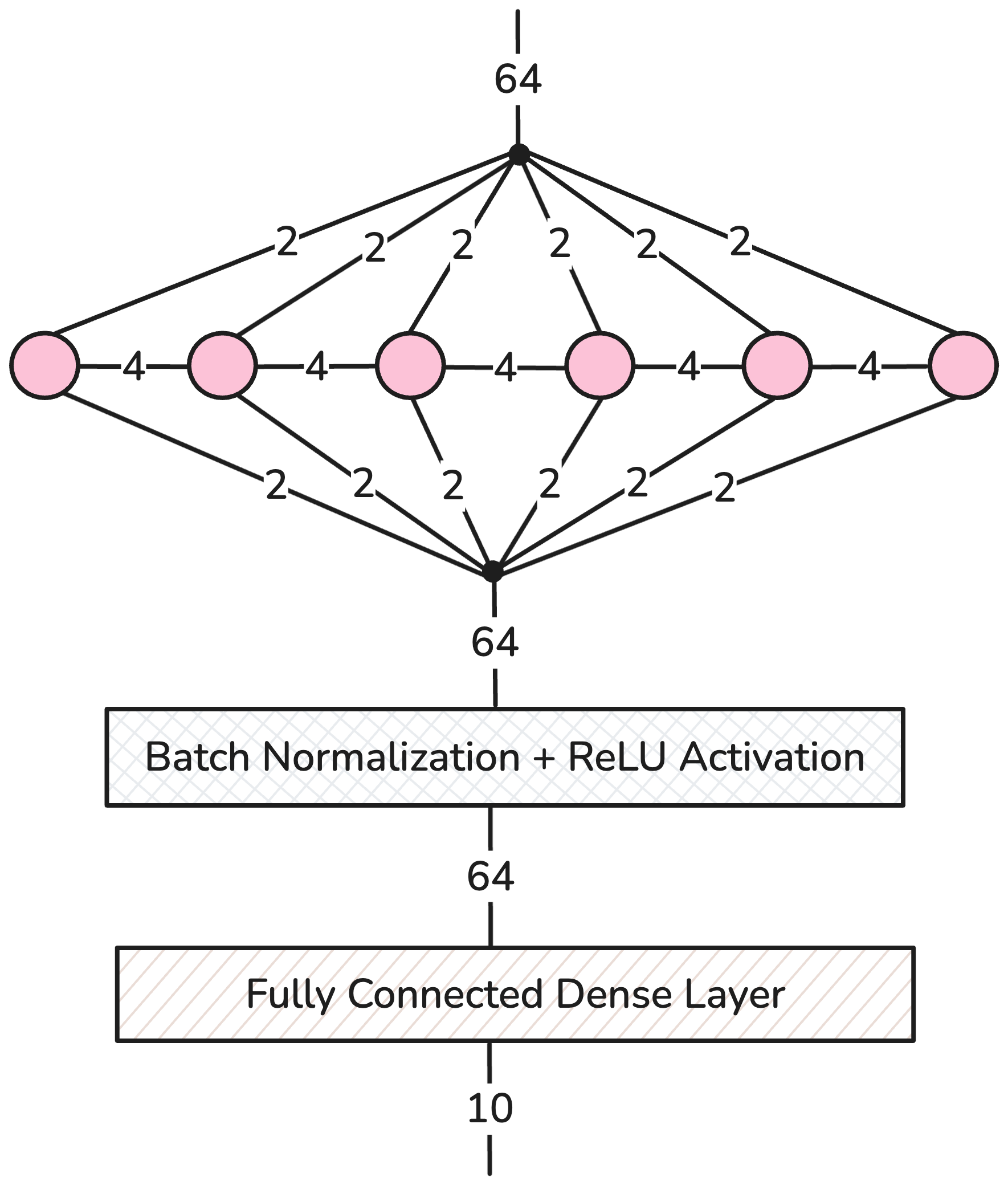}
    \caption{[Color online] A simple classical tensorized neural network (TNN) used in the numerical experiments for variational disentanglers. The network begins with an MPO layer composed of six tensors (shaded circles), with all index dimensions indicated. Black dots represent \texttt{reshape()} operations: the top dot reshapes an input vector of size 64 into a 6-index tensor (each index of dimension 2), while the bottom dot reshapes a 6-index tensor back into a size-64 vector. The second layer applies standard batch normalization followed by a ReLU activation. The final layer is a fully connected dense layer represented by a $64 \times 10$ matrix.}
    \label{fig:variational_experiments}
\end{figure}

\subsection{Numerical results} 

    



\begin{figure*}[ht!]
  \centering
  \subfloat{\includegraphics[width=.48\textwidth]{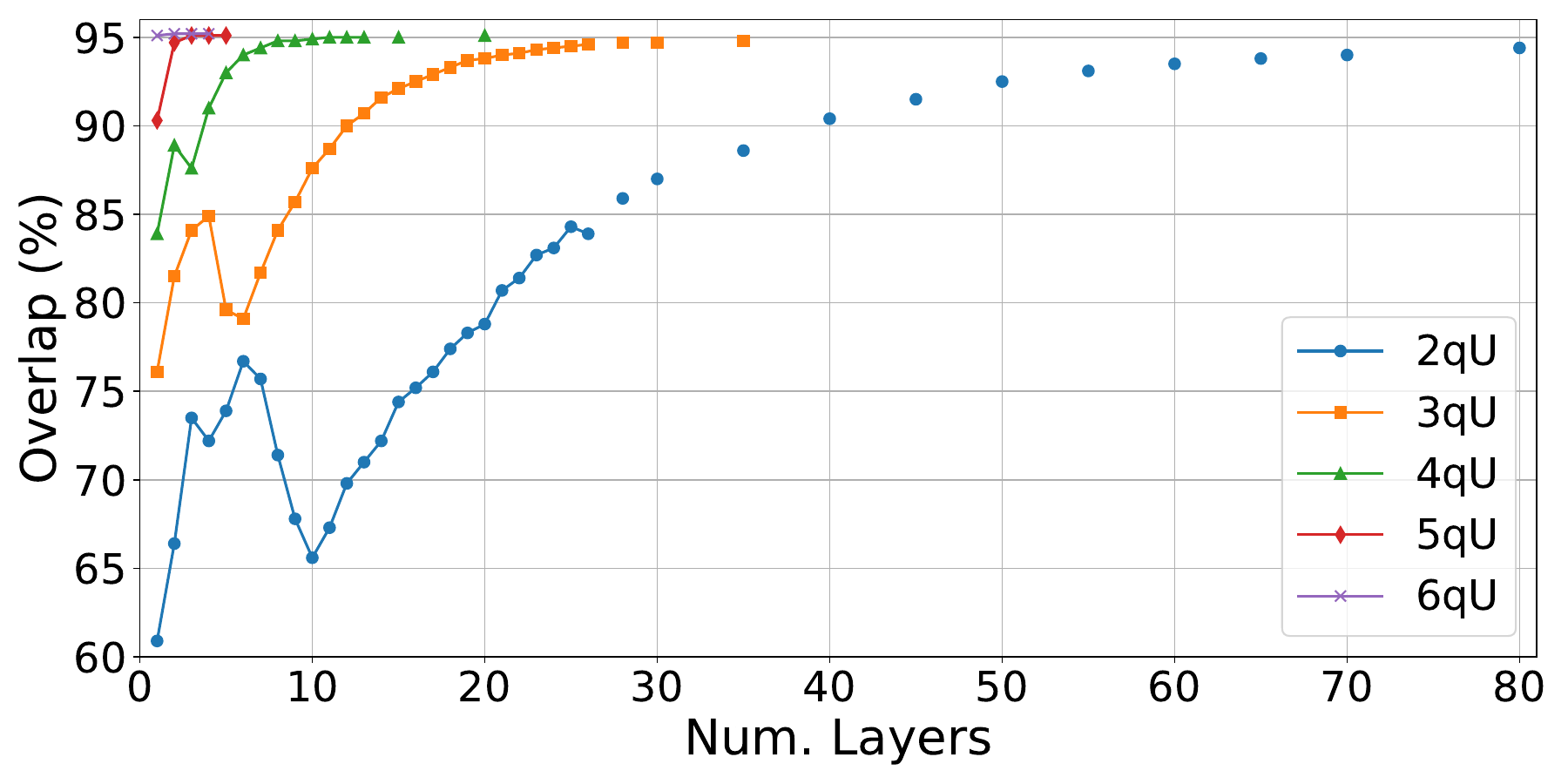}}\hfill
  \subfloat{\includegraphics[width=.48\textwidth]{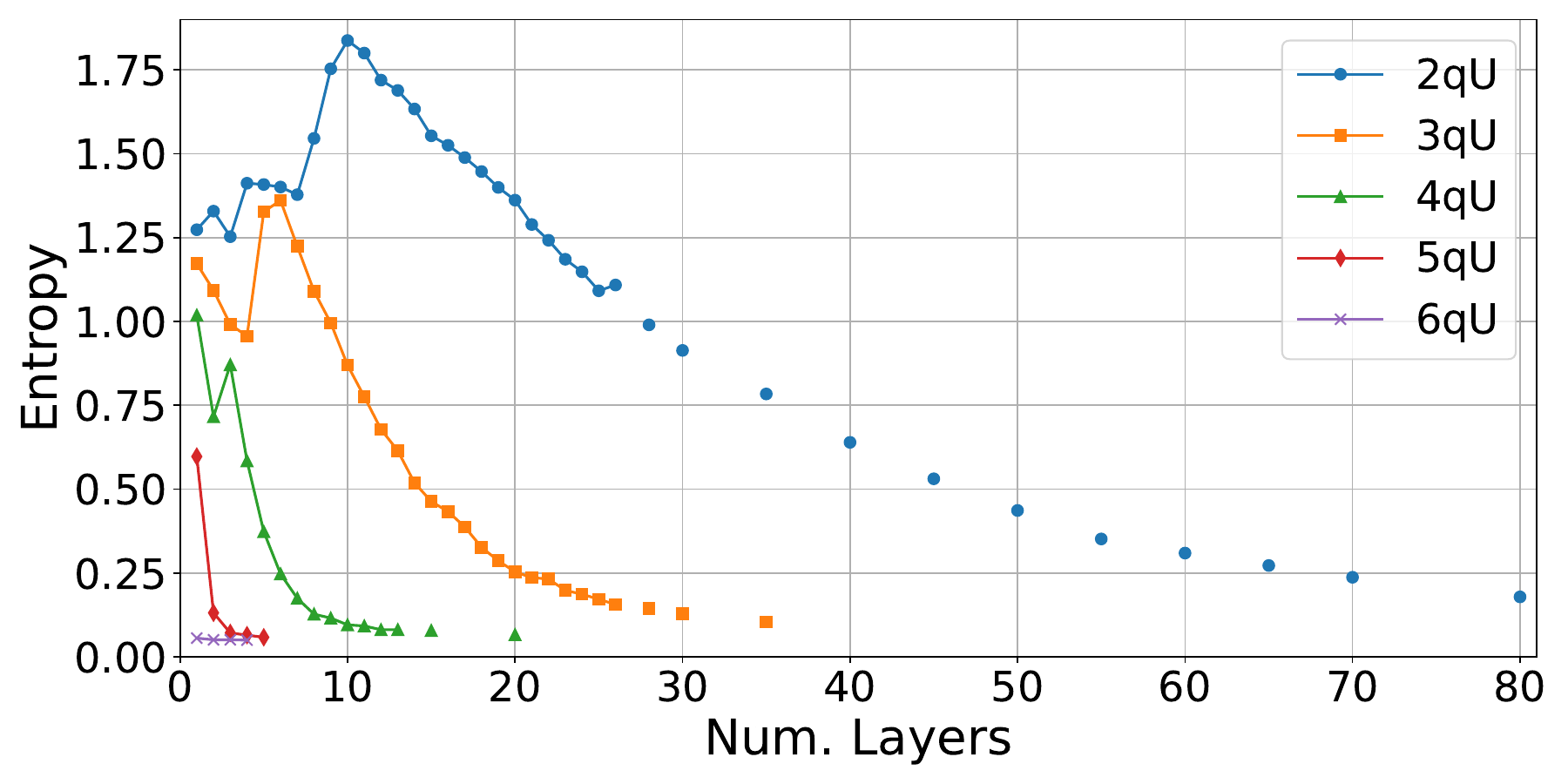}}\\[0.5em]
  \subfloat{\includegraphics[width=.48\textwidth]{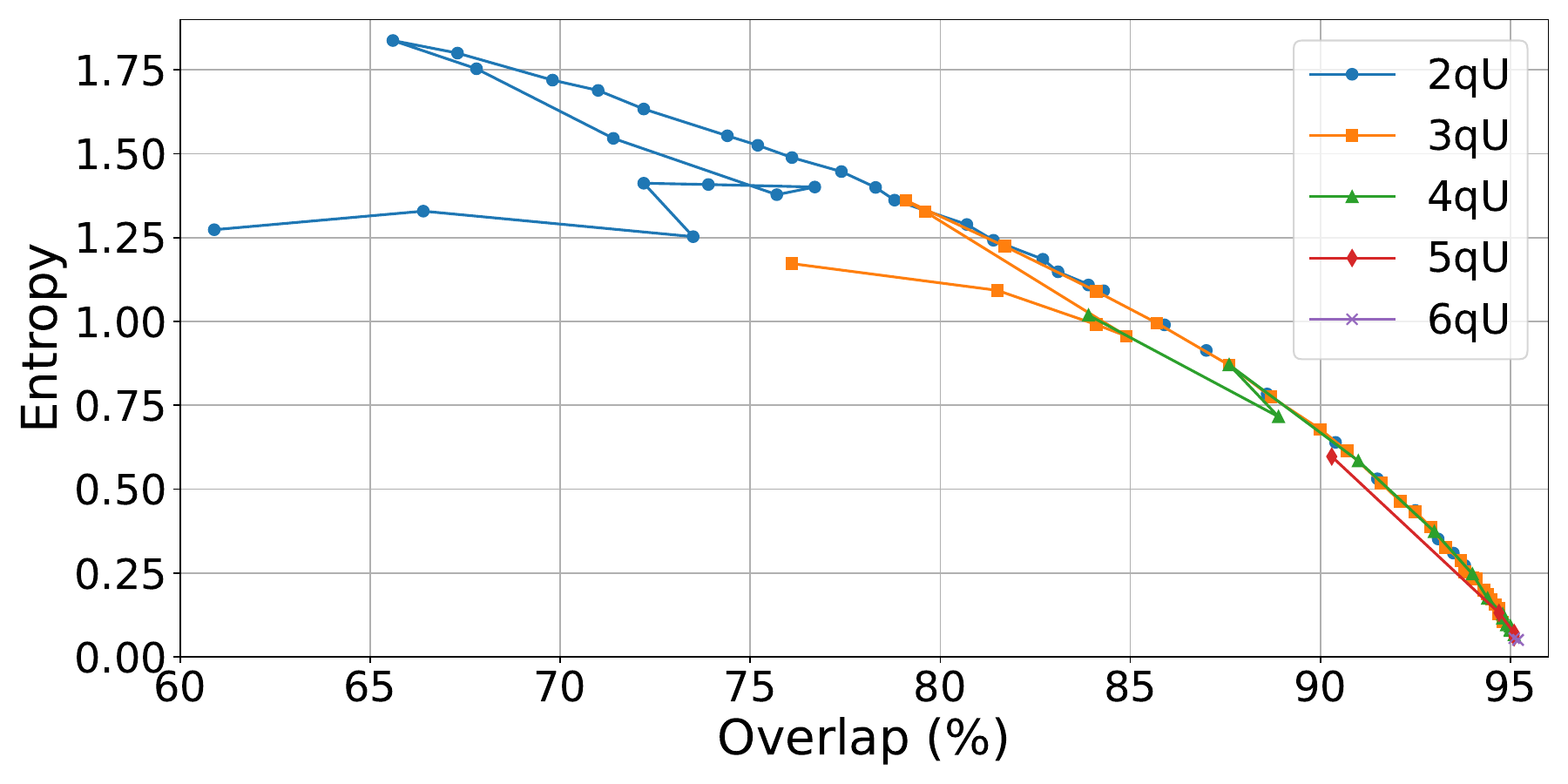}}\hfill
  \subfloat{\includegraphics[width=.48\textwidth]{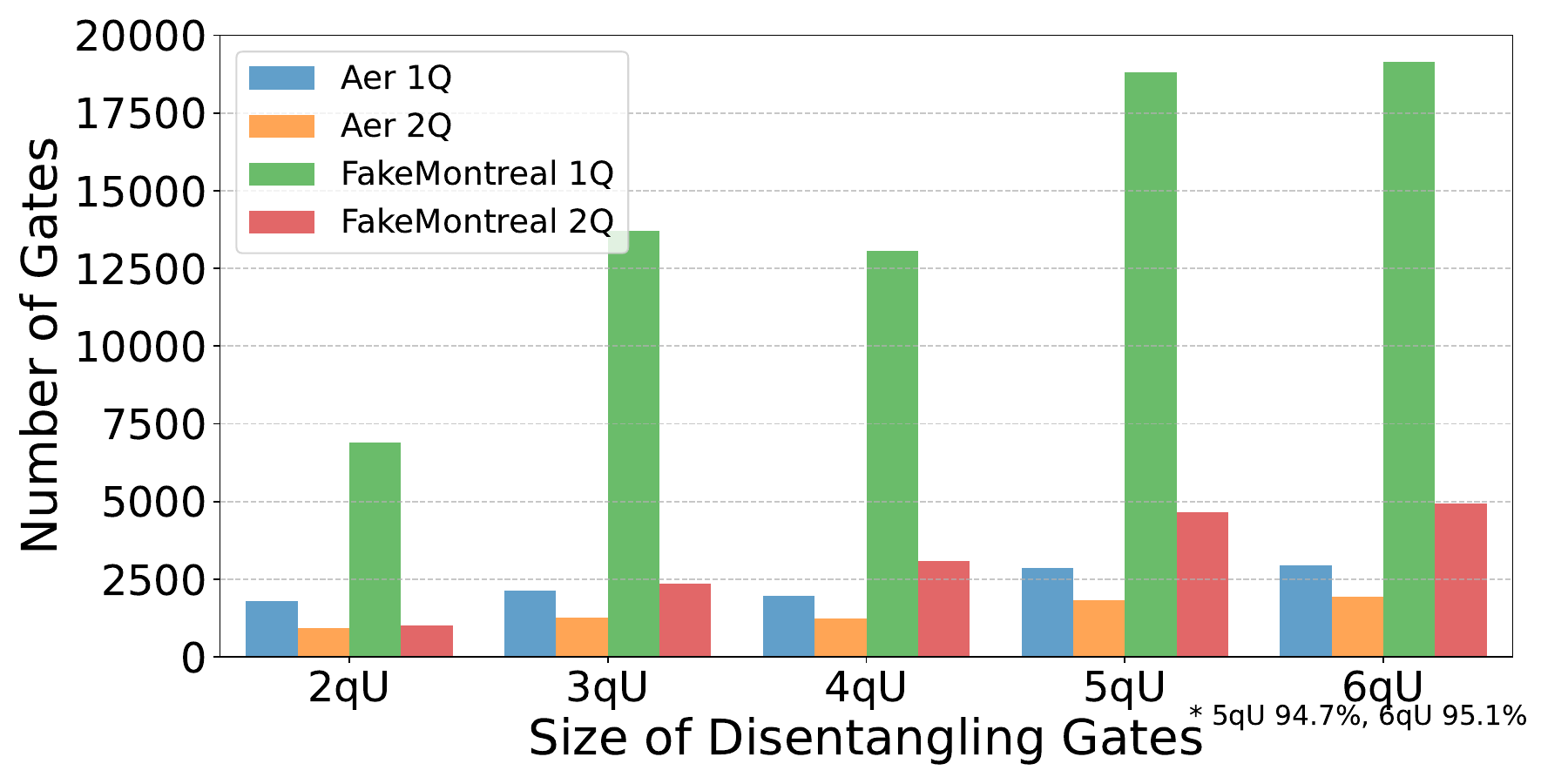}}
  \caption{(Top Left) Overlap, Eq.~\ref{eq:var_loss}, between disentangled MPO and the rank-1 target, as a function of the number of disentangling layers. Deeper circuits improve disentangling. (Top Right) Average Bond Entanglement entropy $S_{\mbox{\tiny avg}}$, Eq.~\ref{eq:avgEE}, of the MPO versus number of disentangling layers. (Bottom Left) Relationship between MPO overlap and entanglement entropy. Higher fidelity correlates with reduced entropy. (Bottom Right) Total gates required to reach $\sim94\%$ overlap. Two qubit unitary (2qU) circuits minimize hardware overhead, making them favorable for NISQ platforms.}
  \label{fig:mpo_results}
\end{figure*}

To practically demonstrate the variational optimization of disentanglers, we begin with a simple proof-of-principle experiment. In this setting, we skip the MPO compression stage---necessary when working with pre-trained dense networks---and instead train a compact tensor network neural network (TNN) from scratch for MNIST classification. Our lightweight model, which attains a test accuracy of 92.6\%, serves as a convenient proxy for the compressed network in our classical-quantum workflow, allowing us to focus directly on the disentangler optimization step.

For this experiment, we downscaled the original MNIST images from \(28 \times 28\) pixels to \(8 \times 8\), resulting in input vectors with 64 components. The TNN architecture comprised: (i) a single MPO layer consisting of six tensors, each with input and output dimension 2 (corresponding to 6 qubit wires, since \(2^6 = 64\)) and all bond dimensions set to 2; (ii) a ReLU (rectified linear unit) activation layer; and (iii) a dense output layer of dimension \(64 \times 10\). This architecture is depicted in Fig.~\ref{fig:variational_experiments}.

Next, we applied the variational disentangling procedure to the trained MPO layer. For this demonstration, we fixed\footnote{More generally, one could jointly optimize both the disentanglers and the compressed MPO tensors. The latter can be updated in a similar fashion, for example via an SVD of the MPO tensor environments appearing inside the trace in the overlap expression~(\ref{eq:var_loss}).} the compressed MPO \(M'_{\chi'}\) as follows: we first brought the input MPO \(M_\chi\) into canonical form, then truncated all but the largest singular value at each bond, resulting in a compressed bond dimension \(\chi' = 1\). This rank-1 MPO served as the target for the disentangling optimization, in which we maximized the overlap~(\ref{eq:var_loss}) to best approximate \(M'\).

We considered disentangling circuits with brickwall geometry composed exclusively of \(n\)-body gates, where \(n = 2, 3, 4, 5, 6\). In Table~\ref{tab:unitary_overlap}, we list the overlap between the target, completely disentangled (\(\chi' = 1\)) MPO, and the input MPO after applying only the first layer of disentanglers on each side. We see that increasing the gate size consistently boosts the overlap---from 62\% for 2-qubit gates to 96\% for 6-qubit gates---demonstrating that larger gates can suppress entanglement more effectively in a single step. This benefit, however, comes at a steep cost: the total gate count grows rapidly, and the number of hardware-compiled gates (evaluated using the FakeMontreal-v2 backend) increases by more than two orders of magnitude when moving from \(n = 2\) to \(n = 6\). This highlights a clear trade-off between circuit efficiency and entanglement suppression.


\begin{table}[htbp]
\centering
\footnotesize
\caption{Effect of unitary size on MPO overlap and gate complexity. Hardware-compilable gate counts (FakeMontreal backend) are higher for larger unitaries.}
\label{tab:unitary_overlap}
\begin{tabular}{|c|c|c|c|}
\hline
\shortstack{\textbf{Gate} \\ \textbf{Size}} & \shortstack{\textbf{Gate Count} \\ (1Q / 2Q)} & \shortstack{\textbf{RealHW Gates} \\ (1Q / 2Q)} & \shortstack{\textbf{Overlap} \\ (\%)} \\ \hline
2qU & 8 / 3 & 32 / 3 & 62 \\ \hline
3qU & 32 / 19 & 174 / 44 & 72 \\ \hline
4qU & 164 / 100 & 1062 / 249 & 81 \\ \hline
5qU & 693 / 442 & 4481 / 1125 & 93 \\ \hline
6qU & 2893 / 1868 & 18481 / 4799 & 96 \\ \hline
\end{tabular}
\end{table}

Next, we investigated the impact of increasing circuit depth. As shown in Fig.~\ref{fig:mpo_results} (top left), deeper circuits yield progressively more faithful reconstructions of the target MPO at large depth. We found an anomalous dip in overlap between beyond five and ten layers, which diminishes as the gates become more non-local. At present, we do not have an explanation for this behavior.

We also quantified compression quality by computing the \emph{average bond entanglement entropy} of the disentangled MPO. Suppose the MPO has \(b\) bond indices. For each bond \(k = 1, 2, \ldots, b\), we reshape the MPO into a matrix \(M_k\) by grouping all open indices to the left of bond \(k\) into a single input index and all open indices to the right into a single output index. Let \(\lambda_k\) denote the diagonal matrix of singular values of \(M_k\). The bond entanglement entropy is then given by  
\begin{equation}\label{eq:avgEE}
S_k = -\mathrm{Tr}(\hat{\lambda}_k^2 \log \hat{\lambda}_k^2), \quad  
S_{\mathrm{avg}} \equiv \frac{1}{b} \sum_{k=1}^b S_k,
\end{equation}  
where \(\hat{\lambda}_k = \lambda_k / \sqrt{\mathrm{Tr}(\lambda_k^2)}\) are the normalized singular values. Interpreting the MPO as a pure \(2(b+1)\)-body quantum state, \(S_k\) quantifies the entanglement across the bipartition defined by bond \(k\). The maximum value of $S_{\mathrm{avg}}$ is
\begin{equation}\label{eq:max}
S^{\mathrm{max}}_{\mathrm{avg}} = \frac{1}{b} \sum_{k=1}^{b} \mathrm{log}~ \chi_k
\end{equation}
where $\chi_1, \chi_2, \ldots, \chi_b$ are the MPO bond dimensions. This follows from the fact that the maximum possible Schmidt rank for each MPO bond is equal to its corresponding bond dimension.

Figure~\ref{fig:mpo_results} (top right) shows a consistent decrease in entropy with increasing disentangling circuit depth, indicating effective disentangling. This is further corroborated in Fig.~\ref{fig:mpo_results} (bottom left), where overlap fidelity is inversely correlated with the final entropy---reinforcing the interpretation that the circuits genuinely disentangle the MPO.

From a practical deployment standpoint, disentanglers must be both expressive and hardware-efficient. Figure~\ref{fig:mpo_results} (bottom right) compares the number of quantum gates needed to achieve approximately 94\% MPO overlap across different unitary sizes. While five and six qubit gate circuits achieve high accuracy, they incur a prohibitive gate cost. Two-qubit gates are the most efficient, striking a balance between expressivity and NISQ suitability. Strategies for improving the corresponding overlaps in these cases are discussed in Sec.~\ref{sec:discussion}.

We also benchmarked the test accuracy of the various models that we have considered above. The results are listed in Table~\ref{results_table}. Disentangled MPOs constructed from six qubit gates maintain baseline accuracy despite compression. In these simulations, 6qU disentanglers were chosen for convenience; this choice affects the residual entanglement structure but not the classification accuracy, as disentanglers in this first stage only reduce entanglement and are not yet optimized to improve accuracy.

We note that adding trainable six qubit variational gates before and after the MPO further improves accuracy to 94.7\%, while their quantum circuit (QC) counterparts achieve 93.6\%. This points to another opportunity for leveraging quantum computing to boost the model's accuracy---by adding new layers to the network that run on the quantum. Although the unitary blocks are mathematically identical in both cases, the quantum-simulated pipeline requires explicit state preparation and tomography of the output. These steps introduce small reconstruction errors even on the noise-free AER backend, which we believe account for the observed $\sim$1\% drop in accuracy compared to the purely classical implementation.

\begin{table}[ht!]
\centering
\caption{Classification test accuracy for different models with and without variationally optimized disentanglers. Below, classical disentanglers correspond to unitary gates that are run as a custom layer inside a classical neural network (custom neural networks layers are discussed in Sec.~\ref{sec:implicit}), while quantum disentanglers are run as simulated quantum circuits via Qiskit on the noise-free AER backend. }
\label{results_table}
\begin{tabular}{|c|c|}
\hline
\textbf{Model}             & \textbf{Accuracy (\%)} \\ \hline
Original TNN (Fig.~\ref{fig:variational_experiments}) & 92.6 (baseline)                \\ \hline
\shortstack{\\Disentangled MPO + \\ gates (classical)}        & 92.6                   \\ \hline
\shortstack{\\Disentangled MPO + \\ gates (quantum)}           & 92.4                   \\ \hline
\shortstack{\\Disentangled MPO + \\  gates + \\ 6-body unitary (classical)}         & 94.7 (+2.1)            \\ \hline
\shortstack{\\Disentangled MPO + \\ gates + \\ 6-qubit unitary (quantum)}         & 93.6 (+1.2)            \\ \hline
\end{tabular}
\end{table}

The Qiskit quantum simulations summarized in the table were performed as follows (boldface denotes Qiskit modules and routines):
\begin{enumerate}
\item Normalize and encode the input data using \texttt{Amplitude} or \texttt{MPS Encoding}.  
\item Add unitaries to the quantum circuit via the \texttt{Operator} and \texttt{Unitary} methods.  
\item Prepare a \texttt{StateTomography} experiment using \texttt{qiskit\_experiments.library}.  
\item Extract the statevector from the resulting density matrix.  
\item Apply the initial normalization and contract with the reduced MPO.  
\item Repeat the procedure with unitaries applied to the other side of the MPO.  
\end{enumerate}

Finally, we highlight two empirical observations that warrant further investigation but lie beyond the scope of this work. First, the disentangling algorithm described above converged more rapidly and achieved higher overlaps than a gradient-descent optimization of unitaries aimed at minimizing the loss in Eq.~(\ref{eq:var_loss})\footnote{This differs from our second approach for finding disentanglers, described in Sec.~\ref{sec:implicit}. Here, gradient descent was used to optimize unitaries that \emph{explicitly} disentangle the MPO by maximizing the overlap (\ref{eq:var_loss}), as in the variational procedure, whereas Sec.~\ref{sec:implicit} discusses a gradient-descent approach for determining \emph{implicitly} disentangling gates.}. Second, disentangling was consistently more effective and efficient for MPOs composed of positive-valued tensors (all entries positive) than for MPOs with mixed-sign entries. This may be related to recent results indicating that the computational difficulty of contracting a tensor network is influenced by the sign structure of its entries~\cite{tnsignproblem, tnpositivebias}.

\section{Implicit disentanglers via gradient descent}\label{sec:implicit}
The disentanglers can also be obtained via gradient descent during the ``healing'' phase of the model, which takes place in the compression step. In this approach, we replace the dense weight inside the network \(\mathcal{N}\) with the product \(\mathcal{Q}_L M'_{\chi'} \mathcal{Q}_R\), and jointly optimize the disentangling gates in the unitary circuits \(\mathcal{Q}_L\) and \(\mathcal{Q}_R\) together with the ``disentangled'' MPO \(M'_{\chi'}\). 

The optimization here minimizes the model's loss---defined here as the standard cross-entropy loss for image classification---rather than explicitly disentangling the MPOs by maximizing the overlap (\ref{eq:var_loss}). In other words, the gates are tuned to improve the overall model performance (or equivalently, maximize its training accuracy), not to directly suppress entanglement. We nevertheless refer to them as ``disentanglers'' since, together with the compact MPO---typically exhibiting an average bond entanglement entropy (\ref{eq:avgEE}) lower than both the original MPO and the maximum entropy (\ref{eq:max}) allowed by the compressed bond dimensions---they provide an accurate replacement for the original MPOs. Notably, even 1-body unitaries can serve effectively as implicit disentanglers when embedded inside a scaffold of fixed CNOT gates. In practice, we find that this replacement doesn't usually degrade the model accuracy without substantially increasing the model size. In this implicit setting, the trade-off of interest is therefore between (i) increasing the MPO bond dimension to recover accuracy and (ii) keeping the MPO bond dimension fixed while increasing circuit depth to recover accuracy.

To perform this optimization, we implemented a custom PyTorch module representing an abstract layer composed of the quantum circuit \(\mathcal{Q}_L\), a compressed MPO \(M'_{\chi'}\), and the quantum circuit \(\mathcal{Q}_R\). Specifically, we derived a custom class from PyTorch's \texttt{nn.Module} base class to store lists of unitary gates for \(\mathcal{Q}_L\) and \(\mathcal{Q}_R\), along with a list of tensors forming the MPO \(M'_{\chi'}\). We then overrode the \texttt{forward()} method to define the data flow through this custom layer.\footnote{See, for example, the simple PyTorch tutorial: https://docs.pytorch.org/tutorials/beginner/nn\_tutorial.html}

Each unitary gate \(g\) was parameterized as \(g = e^t\), where \(t\) is a randomly initialized lower-triangular matrix. As noted earlier, we restricted our focus to real orthogonal matrices (for implementation convenience and does not limit the generality of the framework), which is enforced by constraining \(t\) to be real and lower-triangular.\footnote{This parameterization also constrains the optimization to orthogonal matrices with determinant equal to 1; however, this restriction did not seem to adversely impact the model accuracy.} 

The \texttt{forward()} method of our custom module multiplies the input batch of activations, represented as a matrix \(X\), with the circuits and MPO to return the output activations $Y$ obtained via a sequence of matrix multiplications,
\begin{equation}\label{eq:forward}
Y = X \mathcal{Q}_L M'_{\chi'} \mathcal{Q}_R.
\end{equation}
Here, by slight abuse of notation, \(\mathcal{Q}_{L,R}\) also denotes the matrix obtained by contracting all gates in the corresponding circuit. In practice, the total multiplication (\ref{eq:forward}) was performed using PyTorch's \texttt{einsum()} function, following an optimal contraction order determined by the built-in \texttt{opt\_einsum()} utility.  

As in the variational optimization of the disentangler, for very deep circuits the cost of the tensor network contraction (\ref{eq:forward}) can be significantly reduced by using established contraction schemes \cite{tncontractionbook, tncontractiongray}. When the input activations \(X\) are represented as a large matrix, they can first be decomposed into a low-rank tensor network prior to performing the contraction (\ref{eq:forward}). However, because the circuits in our experiments were relatively shallow, the direct \texttt{einsum()}-based contraction was sufficient.

During training of our custom models with automatic differentiation in PyTorch, we observed vanishing gradients, especially as the depth of the circuits \(\mathcal{Q}_L\) and \(\mathcal{Q}_R\) increased. We were able to substantially improve model performance by inserting nonlinearities between gates. This effectively segments the circuits \(\mathcal{Q}_L\) and \(\mathcal{Q}_R\) into subcircuits that must be installed separately on the quantum computer, while the nonlinear layers remain on the classical computer.


This presents the principal challenge of training implicitly disentangling circuits via gradient descent: deep disentangling circuits may need to be partitioned into shallower subcircuits, potentially diluting the quantum computational advantage due to the overhead of repeatedly preparing activations on the quantum device, performing measurements, and reconstructing output activations via tomography.



\subsection{Numerical results} 

\begin{figure}[t]
    \centering
    \includegraphics[width=\linewidth]{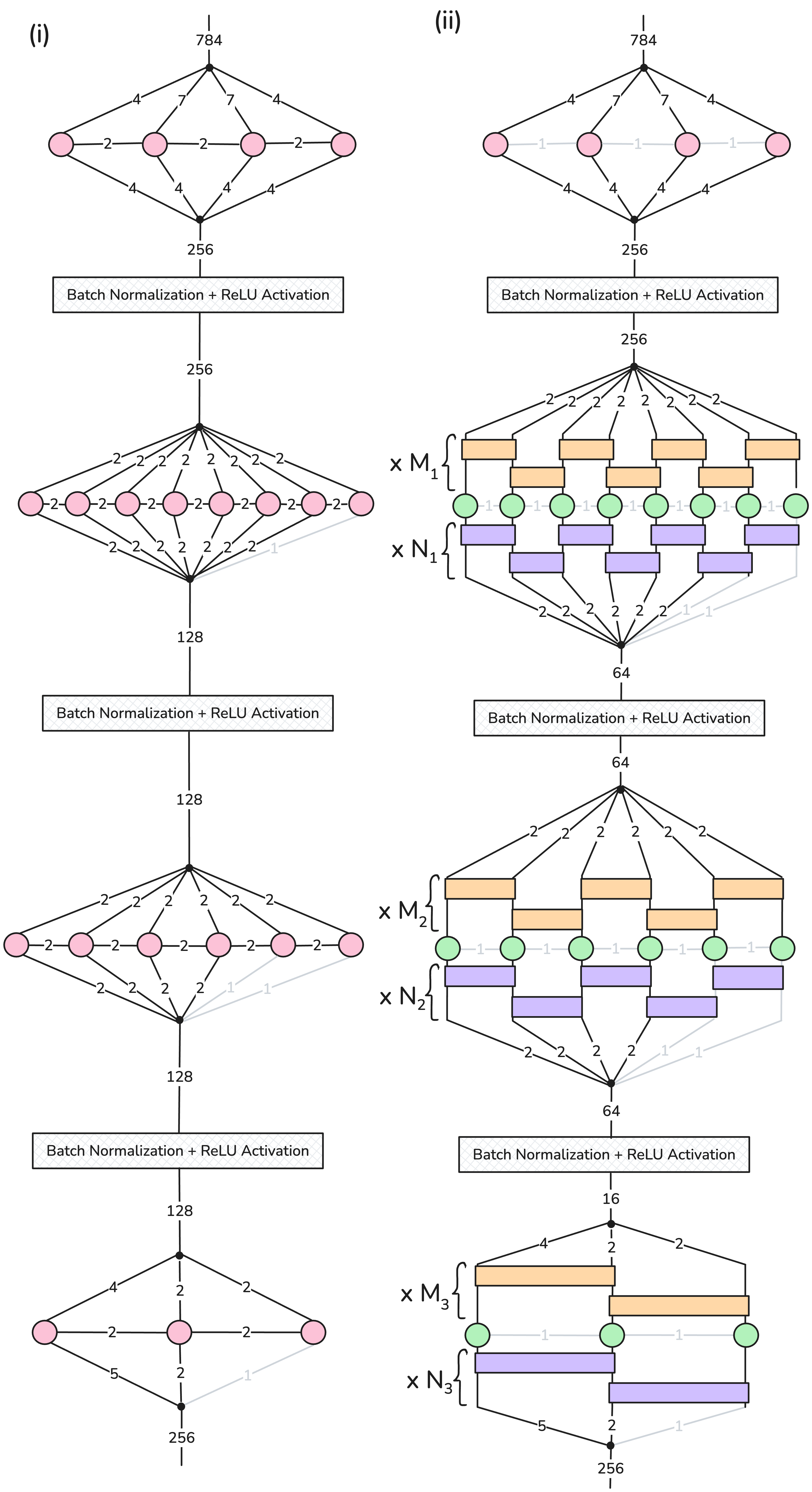}
    \caption{[Color online] Neural network architectures used in the MNIST experiments. Trivial indices having dimension equal to 1 are shown in light gray. (i) The original TNN trained for MNIST classification, comprising four MPO layers (pink) interleaved with batch normalization and ReLU activation functions. (ii) The schematic of the hybrid classical-quantum neural network, featuring completely disentangled MPOs (green) augmented with multiple layers of disentanglers.}
    \label{fig:MNIST_experiments}
\end{figure}

\begin{table}[t]
\caption{Classification test accuracy of different models with and without gradient-descent optimized implicit disentanglers for MNIST classification. The disentangled TNN below is the one depicted in Fig.~\ref{fig:MNIST_experiments}(ii) but without any disentanglers (setting $M_i = N_i= 0$, for all $i$). One-body gates are not shown in the figures. Best model is highlighted in bold font.}
\centering

\label{table:implicit_mnist}
\begin{tabular}{|l|c|c|}
\hline
\textbf{Model}       & \textbf{Params}      & \textbf{Accuracy (\%)} \\ \hline
Original TNN (Fig.~\ref{fig:MNIST_experiments}(i))       &1008        & 94.47 (baseline)                \\ \hline
Disentangled TNN         &854  & 90.31                  \\ \hline
\shortstack{\\Disentangled TNN + \\only CNOTs}   &854       & 91.58                  \\ \hline
\shortstack{\\Disentangled TNN + \\CNOTs + 1-body gates}   &876     & 92.78            \\ \hline
\shortstack{\\Disentangled TNN + \\CNOTs + 2-body gates}  &963        & 93.88            \\ \hline
\shortstack{\\Disentangled TNN + \\CNOTs + 1\&2-body gates}  &930        & 93.51           \\ \hline
\shortstack{\\ \textbf{Disentangled TNN +} \\ \textbf{CNOTs + 1\&2-b gates}}  & \textbf{1029}        & \textbf{94.66}            \\ \hline 
\end{tabular}
\end{table}

\textbf{MNIST classification.} The results of our experiments are listed in Table \ref{table:implicit_mnist}. We first trained the TNN depicted in Fig.~\ref{fig:MNIST_experiments}(i) for MNIST classification obtaining a baseline accuracy of 94.47\%. (We once again bypassed the MPO compression step as we could easily train a tensorized version of the network.) We then consider a target disentangled TNN with the bond dimensions depicted in Fig.~\ref{fig:MNIST_experiments}(ii) without any gates, namely, setting number of layers $M_i = N_i= 0$, for all $i$. After now introducing some 1-body gates (not shown in the figure) and 2-body gates, we perform the gradient descent optimization described above. We also fixed some of the gates to be CNOT gates. The details of the circuit layers inside our MNIST models are listed in Table \ref{tab:circuits_MNIST}. 

Remarkably, including only fixed CNOT gates (kept constant during optimization) led to a slight increase in model accuracy. Introducing trainable one-body and two-body gates separately yielded further improvements, and combining both one-body and two-body trainable gates allowed us to match the baseline accuracy.



\begin{table}[ht]
\centering
\caption{Details of the circuit layers used in the MNIST models listed in Table \ref{table:implicit_mnist}. Here, 1b and 2b denote trainable one-body and two-body gates, respectively, while CNOT gates remain fixed during training. The values $M_i$ and $N_i$ indicate the number of layers in the circuits preceding and following the MPO layers in the model (see Fig.~\ref{fig:MNIST_experiments}(ii)).}
\setlength{\tabcolsep}{3pt} 
\renewcommand{\arraystretch}{1.1} 
\begin{tabular}{l|c|c}
\hline
\textbf{Model} & \textbf{\# Layers} & \textbf{Gate Types} \\
\hline
\multirow{3}{*}{Only CNOTs} 
  & $M_1=1; N_1=2$  & All CNOTs \\ \cline{2-3}
  & $M_2=6; N_2=2$  & All CNOTs     \\ \cline{2-3}
  & $M_3=N_3=1$  & All CNOTs       \\
\hline
\multirow{3}{*}{CNOTs + 1b} 
  & $M_1=1; N_1=2$  & All CNOTs   \\ \cline{2-3}
  & $M_2=8; N_2=1$  & \shortstack{1b + CNOTs + 1b \\+ 5 $\times$ CNOTs; CNOTs} \\ \cline{2-3}
  & $M_3=N_3=1$  & 1b; CNOTs \\
\hline
\multirow{3}{*}{CNOTs + 2b} 
  & $M_1=1; N_1=2$  & CNOTs; CNOTs + 2b \\ \cline{2-3}
  & $M_2=6; N_2=2$  & \shortstack{odd layers = CNOTs \\ even layers = 2b}     \\ \cline{2-3}
  & $M_3=N_3=1$  & \shortstack{odd layers = CNOTs \\ even layers = 2b}      \\
\hline
\multirow{3}{*}{CNOTs + 1\&2-b} 
  & $M_1=1; N_1=2$  & CNOTs; CNOTs + 2b \\ \cline{2-3}
  & $M_2=6; N_2=2$  & \shortstack{odd layers = CNOTs \\ even layers = 2b}     \\ \cline{2-3}
  & $M_3=N_3=1$  & \shortstack{CNOTs + 2b + \\ CNOTs + 2b}
  \end{tabular}
\label{tab:circuits_MNIST}
\end{table}






\begin{figure}[t]
    \centering
    \includegraphics[width=0.95\linewidth]{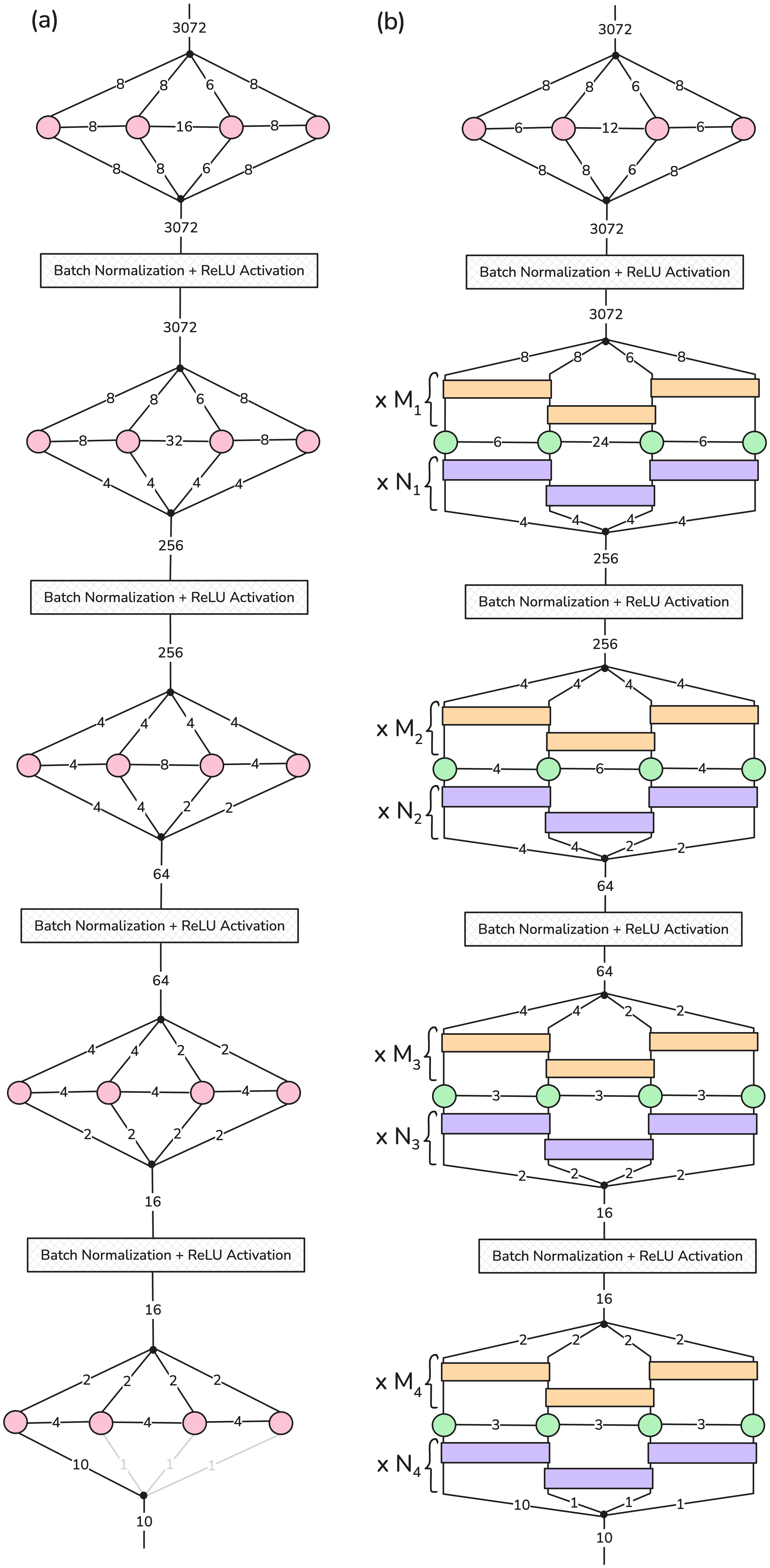}
    \caption{[Color online] Neural network architectures used in the CIFAR experiments. (i) The original TNN trained for MNIST classification, comprising four MPO layers (pink) interleaved with batch normalization and ReLU activation functions. (ii) The schematic of the hybrid classical-quantum neural network, featuring disentangled MPOs (green) augmented with multiple layers of disentanglers.}
    \label{fig:CIFAR_experiments}
\end{figure}

\begin{table}[t]
\caption{Classification test accuracy of different models with and without gradient-descent optimized implicit disentanglers for CIFAR-10 classification. The disentangled TNN below is the one depicted in Fig.~\ref{fig:CIFAR_experiments}(ii) but without any disentanglers (setting $M_i = N_i= 0$, for all $i$). One-body gates are not shown in the figures. Best disentangled model highlighted in bold font.}
\centering
\label{table:implicit_cifar}
\begin{tabular}{|l|c|c|}
\hline
\textbf{Model}       & \textbf{Params}      & \textbf{Accuracy (\%)} \\ \hline
Original TNN (Fig.~\ref{fig:CIFAR_experiments}(i))        &36.7k        & 61.29 (baseline)               \\ \hline
Disentangled TNN         &24.2k  & 58.04                  \\ \hline
\shortstack{\\Disentangled TNN + \\only CNOTs}   &24.2k       & 58.59                \\ \hline
\shortstack{\\Disentangled TNN + \\CNOTs + 1-body gates}   &25.5k     & 59.19           \\ \hline
\shortstack{\\Disentangled TNN + \\CNOTs + 2-body gates}  &24.3k        & 59.90            \\ \hline
\shortstack{\\ \textbf{Disentangled TNN +} \\\textbf{CNOTs + 1\&2-b gates}}  &\textbf{25.3k}        & \textbf{60.74}           \\ \hline
\end{tabular}
\end{table}

\textbf{CIFAR-10 classification.} The results of the CIFAR-10 experiments are presented in Table \ref{table:implicit_cifar}. We began by training a TNN for CIFAR-10 classification with the architecture shown in Fig.~\ref{fig:CIFAR_experiments}(i). The corresponding disentangled TNN, depicted in Fig.~\ref{fig:CIFAR_experiments}(ii), was obtained by setting all $M_i = N_i = 0$. Unlike the MNIST case, we did not fully disentangle the MPOs; instead, we set all MPO bond dimensions to be larger than one but smaller than those in the initial TNN. Once again, adding only CNOT disentanglers resulted in a slight improvement in accuracy. Adding 1-body and 2-body disentanglers also improved accuracy. The details of the circuit layers inside our CIFAR models are listed in Table \ref{tab:circuits_CIFAR}. 

\begin{table}[ht]
\centering
\caption{Details of the circuit layers used in the CIFAR-10 models listed in Table \ref{table:implicit_cifar}. Here, 1b and 2b denote trainable one-body and two-body gates, respectively, while CNOT gates remain fixed during training. The values $M_i$ and $N_i$ indicate the number of layers in the circuits preceding and following the MPO layers in the model (see Fig.~\ref{fig:CIFAR_experiments}(ii)).}
\setlength{\tabcolsep}{3pt} 
\renewcommand{\arraystretch}{1.1} 
\begin{tabular}{l|c|c}
\hline
\textbf{Model} & \textbf{\# Layers} & \textbf{Gate Types} \\
\hline
\multirow{3}{*}{Only CNOTs} 
  & $M_1=N_1=1$  & All CNOTs \\ \cline{2-3}
  & $M_2=3; N_2=4$  & All CNOTs     \\ \cline{2-3}
  & $M_3=4; N_3=3$  & All CNOTs       \\ \cline{2-3}
  & $M_4=N_4=1$  & All CNOTs       \\
\hline
\multirow{3}{*}{CNOTs + 1b} 
  & $M_1=N_1=2$  & All CNOTs   \\ \cline{2-3}
  & $M_2=6; N_2=8$  & \shortstack{odd layers = 1b; \\ even layers = CNOTs \\ remaining CNOTs}  \\ \cline{2-3}
  & $M_3=8; N_3=6$  & CNOTs \\ \cline{2-3}
  & $M_4=N_4=2$  & All CNOTs       \\
\hline
\multirow{3}{*}{CNOTs + 2b} 
  & $M_1=N_1=1$  & CNOTs; 2b \\ \cline{2-3}
  & $M_2=3; N_2=4$  & \shortstack{odd layers = CNOTs \\ even layers = 2b}     \\ \cline{2-3}
  & $M_3=4; N_3=3$  & \shortstack{odd layers = CNOTs \\ even layers = 2b}      \\ \cline{2-3}
  & $M_4=N_4=1$  & 2b + CNOTs       \\
\hline
\multirow{3}{*}{\shortstack{CNOTs + 1b \\ + 2b}}
  & $M_1=N_1=2$  & CNOTs \\ \cline{2-3}
  & $M_2=9; N_2=12$  & \shortstack{1b+CNOTs+1b\\+2b+ CNOTs+2b+\\1b+CNOTs+1b;\\ CNOTs + 1b + CNOTs +\\ 1b + CNOTs + 2b + \\CNOTs + 2b + CNOTs + \\ 1b + CNOTs + 1b + \\CNOTs}     \\ \cline{2-3}
  & $M_3=8; N_3=6$  & All CNOTs \\ \cline{2-3}
  & $M_4=N_4=2$  & All CNOTs
  \end{tabular}
\label{tab:circuits_CIFAR}
\end{table}




The best-performing model was, once again, obtained using a combination of one-body and two-body gates. This model does not yet match the baseline accuracy---owing to its substantially fewer parameters (25.3k) compared to the baseline (36.7k).

We emphasize that the modest parameter reduction in this CIFAR-10 proof-of-concept (36.7k $\to$ 25.3k) is not, by itself, a motivation for deploying quantum hardware. Rather, it illustrates the primary mechanism studied in this work: quantum disentangling circuits can compensate for a reduction in MPO bond dimension, thereby preserving accuracy even when the classical tensor network is deliberately constrained to a smaller bond dimension.

At scale, the relevant constraint is not the absolute number of parameters in these small models, but the classical cost of increasing the MPO bond dimension. Increasing $\chi$ improves the expressivity of an MPO layer, but it also increases the cost of tensor-network contractions and the memory footprint. In contrast, surrounding a fixed-bond-dimension MPO $M'_{\chi'}$ with quantum circuits $\mathcal{Q}_L,\mathcal{Q}_R$ provides an alternative way to recover lost performance without increasing the classical tensor-network complexity.

Beyond this compression role, our results also suggest a secondary implication: circuit layers may serve as an additional expressive resource. In some settings (e.g., the MNIST results in Table~\ref{table:implicit_mnist}), adding trainable quantum gates allows the hybrid model to match or even exceed the original baseline accuracy, indicating that circuit depth can enlarge the effective function class represented by $\mathcal{Q}_L M'_{\chi'} \mathcal{Q}_R$. While this expressive enhancement is not uniformly observed across all experiments, the disentangling framework developed here provides a concrete setting in which such effects can be systematically explored.

We emphasize that batch normalization and ReLU activations are essential in this setting. In particular, when disentangling circuits became too deep, we observed vanishing gradients that led to poor training and higher loss. We substantially improved performance by inserting additional nonlinearities and normalization layers inside the disentangling circuits, beyond those already present in the original TNN. This behavior is reminiscent of the training instabilities reported in deep linear networks (DLNs)-such as plateaus and vanishing gradients with increasing depth \cite{saxe2013exact}-where the absence of nonlinearities fundamentally compromises trainability. With these extra nonlinear layers, our hybrid tensor-network models retained stable training and achieved the accuracies reported in Tables~\ref{table:implicit_mnist} and~\ref{table:implicit_cifar}.

\section{Discussion}  \label{sec:discussion}
The most powerful deep learning models in use today are entirely classical, yet a significant portion of their computational and memory costs is concentrated in a few large linear layers. If even a small fraction of this workload could be offloaded to a quantum computer---without sacrificing accuracy---it could offer a tangible pathway toward near-term quantum utility. In this work, we have explored this possibility by targeting bottleneck layers of a pre-trained neural network, compressing and translating them into quantum circuits, and embedding them within a hybrid classical-quantum inference pipeline.

Our strategy involves first compressing the dense weight matrix \(W\) into a matrix product operator \(M_\chi\), ensuring no significant degradation in model accuracy, and subsequently disentangling \(M_\chi\) into a more compact MPO \(M'_{\chi'}\) with reduced bond dimension \(\chi' < \chi\). In this approximate representation, the disentangling circuits are installed and executed on a quantum processor, while the remainder of the network---including the compressed MPO \(M'_{\chi'}\)---runs on classical hardware. We expect that this two-step approach yields simpler quantum circuits compared to directly approximating \(W\) with a unitary circuit.

The principal challenge to realizing quantum advantage in this framework lies in reducing the size and complexity of the disentangling circuits. While the intermediate MPO representation provides greater control and simplification, a comprehensive exploration of circuit ansatz is needed, including systematic investigation of circuit geometry and gate selection. Notably, the potential of 1-body variational gates combined with fixed, non-variational 2-body gates remains underexplored. Carefully designed circuit ansatz leveraging such gate arrangements could drastically reduce the complexity of hardware-transpiled circuits. Alternatively, one might fix the gate set and optimize solely over circuit geometry---for instance, assessing the extent to which an MPO can be disentangled using only CNOT gates. It would also be interesting to explore disentangling maps composed of more than one Kraus circuit built, corresponding to $k > 1$ in Eq.~\ref{eq:kraus}. 

In terms of computational overhead, the hybrid inference pipeline includes several costs beyond those of a purely classical execution: (i) state preparation to encode classical activations, (ii) execution of the disentangling circuits, and (iii) measurement and classical post-processing to recover output activations. Full state tomography is not expected to scale, and in realistic settings one would estimate only task-relevant observables or use sampling-based reconstruction and structured encodings (e.g., MPS-based state preparation). Therefore, as a proof-of-concept studied here, these quantum input/output costs dominate and preclude any practical quantum speedup. Our aim is not to claim even near-term inference advantage, although future enhancements in quantum technologies may mitigate some of these costs.  Instead, the current benefit of our scheme is that  MPO compression trades dense linear-algebra costs for tensor-network contraction costs that scale polynomially with the MPO bond dimension. Once increasing this bond dimension becomes computationally or memory prohibitive, surrounding the MPO with quantum circuits provides an alternative scaling direction: increasing circuit depth can enhance the effective expressivity of the linear transformation without enlarging the classical tensor network itself.


We remark that the costs associated with measurement and tomography steps in the quantum workflow could be mitigated by either replacing multiple layers by a single effective circuit or by optimally combining (and compiling) circuits from neighboring layers into shorter effective circuits. It may also be possible to reframe measurements and partial tomography as computational \emph{resources} rather than costs, as these operations could be exploited to replace some non-linear transformations within the neural network.

\bibliography{references}

\end{document}